\documentstyle[preprint,aps,epsfig]{revtex}
\preprint{DO/2001/XX/YY\\helsinki preprint no} 
 
\def\be{\begin{equation}}
\def\ee{\end{equation}}

\def\ba{\begin{array}} 
\def\ea{\end{array}} 
\def\bea{\begin{eqnarray}} 
\def\eea{\end{eqnarray}} 
 
\def\l{\left} 
\def\r{\right}

\newcommand{\lsim}{{\;\raise0.3ex\hbox{$<$\kern-0.75em\raise-1.1ex\hbox{$\sim$}} 
\;}} 
\newcommand{\gsim}{{\;\raise0.3ex\hbox{$>$\kern-0.75em\raise-1.1ex\hbox{$\sim$}} 
\;}} 
 
\begin{document} 
\rightline{DO-TH/01/15}
\rightline{HIP-2001-52/TH}

\begin{center}
{\large \bf Revisiting pseudo-Dirac neutrinos}

\medskip

{K. R. S. Balaji\footnote{balaji@zylon.physik.uni-dortmund.de}}

{\it  Institut f\"ur Theoretische Physik, Universit\"at Dortmund,
Otto-Hahn Str.4, 44221 Dortmund, Germany.}

\smallskip

{Anna Kalliom\"aki\footnote{amkallio@pcu.helsinki.fi}}

{\it Theoretical Physics Division, Department of Physical Sciences, University of Helsinki, Finland.}

\smallskip

{Jukka Maalampi\footnote{maalampi@phys.jyu.fi}}

{ \it Department of Physics, University of Jyv\"askyl\"a, Finland}

\smallskip

{\it and}

\smallskip 

{\it Helsinki Institute of Physics, Helsinki, Finland.}  

\date{\today} 
 
%\maketitle 

\end{center}
\noindent
\begin{abstract} 

We study the pseudo-Dirac mixing of left and right-handed neutrinos in the case
where the Majorana masses $M_L$ and $M_R$ are small when compared 
with the Dirac mass, $M_D$. The light Majorana masses could be generated 
by a non-renormalizable operator 
reflecting effects of new physics at some high energy scale. In this
context, we obtain a simple model independent 
closed bound for $M_D$. A phenomenologically consistent scenario is
achieved with $M_L,M_R\simeq 10^{-7}$ eV and $M_D\simeq 10^{-5}-
10^{-4}$ eV. This precludes the possibility of positive mass searches in the
planned future experiments like GENIUS or in tritium decay experiments. If on
the other hand, GENIUS does observe a positive signal for a Majorana mass 
$\geq 10^{-3}$ eV, then with very little fine tuning of neutrino parameters,
the scale of new physics could be in the TeV range, but pseudo-Dirac
scenario in that case is excluded. We briefly discuss the constraints
from cosmology when a fraction of the dark matter is composed of
nearly degenerate neutrinos.

\end{abstract} 
 
\vspace{0.2cm} 
\leftline{PACS numbers: 12.15.Ff, 12.20.Fv, 14.60.Pq, 14.60.St}

\newpage 
 
\section{Introduction} 

%\noindent 

Measurements of the atmospheric neutrino fluxes by the
Super-Kamiokande experiment  
\cite{ref1} and of the solar neutrino fluxes by several experiments \cite{solarexp} have given a 
compelling experimental evidence for neutrino masses, mixing and oscillations. 
The recent results of the SNO experiment \cite{SNO} favour the existence of 
neutrino oscillation 
among active flavours involving $\nu_e$ from the Sun. Upon inclusion of the  
LSND result \cite{LSND}, a simultaneous explanation of both the solar and atmospheric results in terms 
of oscillations would require the existence of at least one sterile neutrino which can oscillate with 
any of the active flavours. There are many analyses
 in the literature where various possible active-sterile neutrino oscillation patterns have been 
studied~\cite{analyses}.   
 
In most analyses, the atmospheric anomaly points for its
solution towards large angle $\nu_{\mu}\to\nu_{\tau}$ or $\nu_{\mu}\to\nu_{s}$ 
oscillations, where $\nu_{s}$ denotes a sterile 
neutrino. Results obtained by CHOOZ reactor based $\bar{\nu}_e$ 
disappearance experiment~\cite{CHOOZ} and later by PaloVerde 
\cite{paloverde} severely constrain $\nu_{\mu}\to\nu_{e}$ oscillations 
for neutrino mass scales relevant for atmospheric neutrinos. 
This is also in agreement with the flat spectrum observed for
the atmospheric {\it e}-like events. In addition, an analysis of the
neutral current data disfavours large transitions involving $\nu_e$ at
the atmospheric scale~\cite{balaji}.
Recently, the Super-Kamiokande collaboration  
argued that the oscillations between active-sterile flavours 
is disfavoured at 3$\sigma$ level~\cite{superkamatm}. It should be
mentioned, however, that this conclusion may depend on how one
analyses the data, and it has been claimed that  a maximal 
$\nu_{\mu}\to\nu_{s}$ oscillation solution to 
the atmospheric neutrino problem is not yet ruled out~\cite{foot}. 
Furthermore, it has been argued that the study of neutral 
current events at Super-Kamiokande, combined with the information
obtained from future long baseline experiments, might not even be
sufficient to decide between active-active and active-sterile oscillation  
solutions~\cite{geiser2}. 
 
The possible role of the active-sterile oscillations in explaining the solar neutrino problem has
 recently got new light from the first SNO results on the charged current rates. The pre-SNO 
situation was such that active-sterile large mixing angle (\mbox{LMA})  
as well as low mass (\mbox{LOW}) solutions were disfavoured whereas 
small mixing angle (\mbox{SMA}), vacuum (\mbox{VAC}) and \mbox{Just-So} 
solutions were well allowed~\cite{Bahcall}. Upon inclusion of the preliminary SNO results, 
within the two flavour analysis, it appears that only the VAC solution gives a good fit to the 
data with best fit point as $\Delta m^2_\odot=1.4\cdot 10^{-10}\;{\rm eV}^2$ and 
$\tan^2\theta_\odot=0.38$ \cite{KrastevSmirnov}. Alternatively, magnetic moment solutions to the 
solar anomaly are also feasible. Such solutions equally involve large active-sterile oscillations 
and are currently not ruled out \cite{akh}.
 
It may of course be that the solar neutrino oscillations follow in reality a more complicated 
pattern than an effective two flavour scenario.  
The SNO and future experiments, especially those which are sensitive 
to both charged and neutral currents (Borexino and KamLAND), are believed to provide a crucial test of
 the existence of  
oscillations to sterile neutrinos of any form.  On the other hand, Barger et al.~\cite{Barger}  
have recently argued that due to the poorly known value of the $^8$B flux normalization,  
even the forthcoming SNO neutral current measurement might not be  
sufficient to determine the sterile neutrino content in the solar neutrino flux.  
 
Thus, given our current understanding and analyses of the neutrino 
data, large active-sterile oscillations may play some role in 
solving the solar and atmospheric neutrino anomaly,  
though it seems to be less probable than active-active solutions.  
Furthermore, a combined analyses of the neutrino data including the LSND result favours
a $2+2$ spectrum which involves the possibility of large active-sterile oscillations either 
in the solar or atmospheric sector \cite{grimus}.
 
   All of the above solutions require neutrinos to posses a small but  
non-vanishing mass. From the theoretical point of view, the seesaw  
mechanism \cite{ref9} offers the simplest and the most natural  
explanation for small neutrino masses. In this mechanism, one assumes the existence of a  
large Majorana mass scale  ($M_R$) for the right-handed neutrino  
($\nu_R$), $M_R \gg M_D ~\mbox{and} ~M_L$. Here, $M_D$ is a Dirac  
mass and $M_L$ is a Majorana mass of the left-handed neutrino 
($\nu_L$), both of which occur in a general Dirac-Majorana mass Lagrangian 
for $\nu_L$.  
Upon diagonalization, the seesaw mass Lagrangian leads to two Majorana neutrinos, one
 with a very small mass ($\sim M_D^2/M_R)$ and another one with a large mass ($\sim M_R$). 
Therefore, the sterile neutrino in this scheme decouples from the low energy world and cannot 
play any role in the oscillation phenomena under discussion. 
  
If, on the other hand, one assumes $M_D \gg M_R,~M_L$,  
the situation is quite different. The resulting mass eigenstates have 
eigenvalues very close to each other, and they have opposite 
CP parities. Hence they can form a pseudo-Dirac 
neutrino~\cite{wolf}. There have been numerous suggestions 
in the literature for pseudo-Dirac neutrinos as solution to the 
neutrino anomalies, where the observed flavour suppression is 
due to a maximal or near to maximal mixing between an active and a  
sterile neutrino~\cite{ref10}.  
 
A relevant question in the pseudo-Dirac scenario is to explain the unorthodoxy in the hierarchy:
$M_D \gg M_R$ which is necessary for sterile neutrinos to be light. In the standard model (SM) the 
Majorana masses $M_L$ and $M_R$ are non-existing due to the conservation of lepton number. Hence 
the origin of these mass terms goes beyond the SM and there could be many sources. One possibility is
that the masses may be provided at the SM level by non-renormalizable effective 
operators of the type $L^2\phi^2/M$ and $\nu_R^2\phi'^2/M'$. Here $L=(\nu_L,l_L)$ is an ordinary 
lepton doublet, $\phi$ and $\phi'$ are Higgs fields, and $M$ and $M'$ are high mass scales derived 
from some beyond-the-SM theory. The masses $M$ and $M'$ are not necessarily connected with the vacuum 
expectation values of the Higgs fields $\phi$ and $\phi'$, so it is possible that both $M_L$ and $M_R$
 are much smaller than $M_D$. In any case, it is known that in a viable model $M_L$ should be 
suppressed so that $M_D \gg M_L$. This is required to avoid a contradiction with the accurately 
determined $\rho$-parameter. It is conceivable to assume that a similar suppression also happens for $M_R$. 
 
A subsequent question is to understand the smallness of $M_D$. A light Dirac mass can be 
either (i) due to a small Yukawa coupling in the mass term $\overline\nu_R\nu_L\phi$ or (ii) just like
in the case of $M_L$ or $M_R$, a  light $M_D$ could be generated by 
a non-renormalizable higher dimensional term \cite{ChangKong}.
 Another possibility is realized in models with large extra spatial dimensions. In such theories,
 the Yukawa coupling of the term $\overline\nu_R\nu_L\phi$ may be suppressed as the right-handed 
neutrino can be most of the time in the bulk outside our four-dimensional brane \cite{extradim}. 
In the following, we assume a small $M_D$ relevant for a pseudo-Dirac mixing without addressing
to its origin. We examine the mixing of $\nu_L$ and $\nu_R$ when the Dirac mass term 
dominates over the Majorana mass terms, i.e. $M_D \gg M_L, M_R$, and discuss the experimental and 
theoretical bounds one can obtain for the mass parameters. This is illustrated for the case of the electron 
neutrino.

 Our paper is organised as follows. In the next Section, we give the basic formalism for 
pseudo-Dirac mixing and by a simple
exercise we show that the effective electron neutrino mass as probed by neutrinoless double beta 
decay experiments is exactly $M_L$. In Section III, we set bounds for the masses, $M_L$ and $M_D$
and derive a closed bound for $M_D$. We also discuss the 
constraints from cosmology when some fraction of the dark matter is composed of nearly degenerate 
neutrinos. Finally, in Section IV, we conclude by summarising the main results of this paper.  
 
\section{The Pseudo-Dirac scenario}

Let us consider the $2\times2$ Dirac-Majorana mass matrix in the 
$(\nu_L, \nu_L^C)$ basis of the form 
\[ 
\cal{M}=\l(\ba{cc} M_L & M_D \\ M_D & M_R \ea \r)~,  
\] 
and assume $M_D \gg M_L, M_R$. 
The mixing angle which diagonalises $\cal{M}$ is easily derived to be 
 
\be 
\tan 2\theta =\frac{2 M_D}{M_R - M_L}~. 
\label{eq1} 
\ee 
% In the limit, $M_D \gg (M_L,~M_R)$  
We get a pseudo-Dirac neutrino pair with mass eigenvalues 
 
\be 
m_{\pm} = \delta~\pm~\bar M ~, 
\label{eq2} 
\ee 
 
\noindent 
where $\bar M =(M_L+M_R)/2$ and  
$\delta=\sqrt{(M_L-M_R)^2 + 4M_D^2}/2 \approx M_D$ . For a nonzero $M_D$  
and $M_L=M_R$, this system corresponds to a  
maximal interlevel mixing of $\pi/4$ between the Majorana pair. If 
$M_D>0$ is assumed, the neutrino mass-squared difference is 
 
\be 
\Delta m^2 =m_+^2-m_-^2=4M_D\cdot M_L~. 
\label{eq3} 
\ee 
If $M_L\neq M_R$, i.e. when the mixing is not maximal, one has 
 
\be 
\Delta m^2 \simeq 2M_D (M_L+M_R)~. 
\label{eq3.1} 
\ee 
 
In the case of $M_D \gg M_L, M_R$, which we are interested in here, the Majorana mass 
parameters $M_L$ and $M_R$ do not contribute substantially to the 
kinematical masses $m_{\pm}$. As a result, the standard mass measurements 
based on particle decays are not sensitive to them but only probe the Dirac 
mass parameter $M_D$. The parameters $M_L$ and $M_R$ can be  
tested in processes where they have a dynamical role. The most 
important process for studying $M_L$ is the neutrinoless double beta 
$(0\nu\beta\beta)$ decay. One can easily see that the effective  
neutrino mass $M_{eff}$ measured in $0\nu\beta\beta$ decay experiments 
is actually $M_L$ and it is independent of $M_D$ and $M_R$. The mass eigenstates, 
$\nu_L^\pm$, can be written in terms of the interaction states, $\nu_L$ and 
$\nu_L^C$, as 
 
\be 
\nu_L^\pm = N_\pm 
\big[2M_D~\nu_L +(M_R-M_L\pm 2\delta)~\nu_L^C\big]~,  
\label{j1} 
\ee 
where
\be 
N_\pm  = \big[2(M_R-M_L)^2 + 8M_D^2 \pm 4(M_R-M_L)\delta\big]^{-1/2} 
\ee 
are normalization factors. In the limit $M_D \gg M_L ~\mbox {and}~M_R$, 
 
\bea 
\nu_L^\pm &\simeq& N_\pm 
\big[2M_D~\nu_L +(M_R-M_L\pm 2M_D)~\nu_L^C\big]~;\nonumber\\ 
N_\pm &\simeq& \frac{1}{2\sqrt{2}M_D}\big[1\mp \epsilon\big]~;~ 
\epsilon=\frac{M_R-M_L}{4M_D}\ll 1~. 
\label{j2} 
\eea 
Therefore, the active neutrino component $\nu_L$ in the mass eigenstates is given by the amplitude 
 
\be 
\langle\nu_L|\nu_L^\pm\rangle =2M_DN_\pm \simeq \frac{1}{\sqrt{2}} 
(1\mp \epsilon)~, 
\label{j3} 
\ee 
 
\noindent 
implying 
 
\be 
\nu_L=\frac{1}{\sqrt{2}}\big[(1-\epsilon)\nu_L^+ +  
(1+\epsilon)\nu_L^- \big]~. 
\label{j4} 
\ee 
 
 The effective electron neutrino mass as 
measured by $0\nu\beta\beta$ decay experiments is then given to be 
 
\be 
M_{eff} = \cos ^2\theta~\eta_+~m_++\sin ^2 \theta~\eta_-~m_- 
=\frac{1}{2}\big[(1-\epsilon)^2 \eta_+m_+  
+ (1+\epsilon)^2 \eta_-m_-\big] \approx M_L~, 
\label{j5} 
\ee 
 
\noindent 
where in $\eta_\pm=\pm 1$ are the Majorana phases of the mass eigenstates. 
It is easy to check that without the 
assumption $M_D\gg M_L, M_R$ one ends up with the exact result $M_{eff}=M_L$.  
 
\section{Bounds for $M_L$ and $M_D$} 
 
\noindent{\it A lower bound for $M_D$.} 
From (\ref{eq3.1}) and (\ref{j5}) a general result follows:
 
\be 
M_{eff} \simeq \frac{\Delta m^2}{2 \beta M_D}~, 
\label{eq4} 
\ee 
 
\noindent 
 where $\beta\equiv M_R/M_L +1 > 1$. The 
most stringent experimental upper bound published by the
Heidelberg-Moscow experiment in \cite{HeiMos} implies $M_{eff} \leq
M^{exp}_{eff}=0.2$ eV (more recently the experiment has quoted the limit
0.34 eV at 90 \% C. L. \cite{klapdor}). Thus, for a 
given $\Delta m^2$, to be 
consistent with $0\nu\beta\beta$ decay results, the Dirac mass $M_D$ must obey the bound   
 
\be 
M_D~\gsim ~ \frac{\Delta m^2}{2 \beta M^{exp}_{eff}}~. 
\label{eq5} 
\ee

\noindent{\it A bound for $M_L$ from unitarity.} 
A Majorana mass $M_L$ of the left-handed neutrino 
reflects physics beyond SM. In its presence the SM should be considered as 
an effective theory. It should be replaced by a more fundamental 
theory at some high energy scale $M_X$, where new 
physics should enter, since otherwise the processes induced by the Majorana 
mass term would spoil the unitarity. One can find an upper 
limit for $M_X$, for example, by studying the high energy behavior of the 
lepton number violating reactions $\nu\nu\to WW$ or $ZZ$, which can occur 
because of the Majorana mass term. The amplitudes of these reactions increase as proportional to the 
center of mass energy, leading to a breakdown of the effective theory at high energies. It was 
recently shown \cite{maltoni} that the most stringent bound for $M_X$ is obtained by  
considering the following linear combination of the zeroth partial wave
 amplitudes: 
$a_0(\frac{1}{2}(\nu_{+}\nu_{+}-\nu_{-}\nu_{-})\to\frac{1}{\sqrt{3}}(W^+W^++Z^0Z 
^0))$, 
where $\nu_{\pm}$ are helicity components of the mass eigenstate neutrino 
$\nu$ and the final state bosons are longitudinally polarized. This amplitude to obey unitarity,
 i.e. $\vert a_0\vert\leq 1/2$, requires~\cite{maltoni} 
 
\be 
M_X\leq \frac{4\pi\langle\phi\rangle^2}{\sqrt{3}M_L}~, 
\label{maltoni} 
\ee 
where $\langle\phi\rangle=174$ GeV is the vev of the ordinary Higgs boson. It 
should be stressed that the Majorana mass $M_L$ appears in this 
formula, not the kinematical mass of the neutrino. At high energies, where neutrinos are 
ultra-relativistic, the kinematical mass of the neutrino is irrelevant. 
 
The condition (\ref{maltoni}) can be used to set an upper limit for 
the Majorana mass $M_L$. If new physics starts to operate at the 
Planck scale $M_{pl}\simeq 1.2 \cdot 10^{19}$ GeV, then 
 
\be 
M_L \leq \frac{4\pi \langle \phi \rangle ^2} 
{\sqrt{3}M_{pl}}\simeq 2\cdot 10^{-5}~\mbox{eV}~. 
\label{r2} 
\ee 
The smaller the scale of the new physics, the less stringent is the bound. 
The $0\nu\beta\beta$ decay to be visible in the planned GENIUS experiment\cite{genius}, i.e.
 $ M_L\gsim 10^{-3}$ eV, would require $M_X\lsim 10^{17}$ GeV.

\noindent 
{\it {A closed bound for $M_D$}}. 
As was originally pointed out by Weinberg \cite{weinberg}, Majorana masses for the left-handed neutrinos
 can be generated by higher dimensional operators of the form 

\be
{\cal L}_5=\frac{f_{\alpha\beta}}{M_X}(L_{i\alpha}^TC^{-1}L_{j\beta}
\phi_{k}\phi_{l}\epsilon_{ik}
\epsilon_{jl})~,
\label{weinberg} 
\ee

where $i,j,k,l$ are $SU(2)_L$ indices, $\alpha,\beta$ are flavour indices, and $M_X$ is the scale of 
new physics. This operator breaks the lepton number explicitly, and after spontaneous symmetry 
breaking it leads to the following Majorana mass (neglecting flavour mixing): 
 
\be 
M_L=f\;\frac{\langle\phi\rangle^2}{M_X}~,
\label{l5} 
\ee 
where $f\lsim {\cal O}(1)$ is a numerical factor. With $M_X\lsim M_{pl}$ this implies 
\be 
M_L\gsim f\;\frac{\langle\phi\rangle^2}{M_{pl}}\simeq 3\cdot 10^{-6}\;
{\rm eV}\cdot f~. 
\label{mlbound}
\ee 
 Therefore, in this scheme we have 
\be 
3\cdot 10^{-6}\;{\rm eV}\cdot f\lsim M_L\lsim 0.2\; {\rm eV}~, 
\label{MLrange} 
\ee 
where the upper bound is due to the $0\nu\beta\beta$ decay results.

By using (\ref{eq4}) one can infer from (\ref{MLrange}) the following closed bound for possible values 
of the Dirac mass $M_D$: 
 
\be 
\frac{\Delta m^2}{0.4\;\beta\;{\rm eV}}\lsim M_D\lsim \frac{\Delta m^2}
{6\cdot 10^{-6}\;f\beta\;{\rm eV}}~. 
\label{MDrange}\ee 
 
Let us now turn to experimental numbers involving the electron neutrino.
 According to the analysis done in \cite{KrastevSmirnov} for 
the solar neutrino problem (that takes into account the recent results of SNO on the $\nu_e$ charged
 current rate), the best fit values for pure vacuum solution ($\nu_e \leftrightarrow \nu_s$) are with  
$\Delta m^2=1.4\cdot 10^{-10}$ eV$^2$ and $\tan^2\theta = 0.38$. 
This does not correspond to a maximal mixing which is the case
in the pseudo-Dirac scenario. However, as can be
seen from the analysis \cite{KrastevSmirnov}, maximal mixing
with $\theta=\pi/4$ is not completely ruled out even though it is less
favoured. To illustrate the situation we set 
$\Delta m^2=1.4\cdot 10^{-10}$ eV$^2$ and $\beta =2$ as reference values which
corresponds to maximal active-sterile mixing in the case $M_L=M_R$. 
With these values, (\ref{MDrange}) gives  the numerical range
\be 
1.8\cdot 10^{-10}\;{\rm eV}\lsim M_D\lsim 2.4\cdot 10^{-5}\;{\rm eV}/f~. 
\label{r1}
\ee 

Comparison with (\ref{MLrange}) 
shows that for the small $\Delta m^2$ of the vacuum solution, the pseudo-Dirac
 requirement 
$M_R, M_L\ll M_D$ leads to a consistent picture only when $M_D$ is in the 
upper end of this range. If we take
 $f=0.1$, then a possible situation could be, e.g., $M_L,M_R\simeq 10^{-7}$ 
eV and 
$M_D\simeq 10^{-5}- 10^{-4}$ eV. In any case, one can conclude that if the
 solar neutrino deficit is due
 to a pure sterile mixing,  $M_L$ is necessarily so small that
 the $0\nu\beta\beta$ 
decay would stay outside the range that the upcoming GENIUS experiment would 
be able to probe. On the 
other hand, the kinematical determination of the electron neutrino mass in 
tritium decay \cite{tritium} would also be
extremely difficult because of the smallness of $M_D$. Nonetheless, the analysis does predict a nonzero
mass value from both these processes and hence the associated scale of new physics\footnote {The scale
can be extracted depending on the specific nature of a model for $M_D$.}.

\noindent{\it TeV scale physics.} 
It follows from (\ref{eq4}) and 
(\ref{l5}), together with the requirement $M_L \ll M_D$, 
that with any natural values of $f$, the  energy scale $M_X$ must be fairly 
close to the Planck scale $M_{pl}$. This can be
illustrated with the following example. If we wanted to have new
physics close the weak scale, e.g. in the TeV scale, it follows from
(\ref{l5}) and the experimental limit $M_L \leq M_{eff}^{exp} =  0.2$ eV
that $f < 10^{-11}$, and further, the requirement $M_L \ll
M_D$ to be satisfied, one  must have $ f < 10^{-16}$. In fact, if
$f$ is $\mathcal{O}$(0.1), the feasible range for new energy scale is
$M_X \gsim 10^{-2}M_{pl}$.  With such high values of $M_X$ 
there is no hope to observe $M_L$ and $M_D$ at least in near future, as 
already mentioned. 

A larger Majorana mass $M_L$  from TeV-scale new physics could be obtained 
in models where there are suitable additional scalars. Within the context of
nonrenormalizable theories, this is feasible if we consider a higher 
dimension operator other than the one suggested in (\ref{weinberg}).
 To illustrate this, we consider the simplest extension to the SM with 
an extra scalar doublet, $\phi^\prime$. In order to avoid the induced flavour 
changing neutral currents, we impose a discrete $Z_2$ symmetry
for the field $\phi^\prime$. In this case, the lowest possible higher 
dimensional operator, which can generate a Majorana mass, is of the type 

\be
{\cal L}_7 = \frac{f^\prime}{M_X^3}(L \phi  \phi^\prime)^2~.
\label{l7}
\ee
A Majorana mass is obtained when the scalars get a vev:
\be
M_L = \frac{f^\prime}{M_X^3}(\langle \phi \rangle \langle \phi^\prime
\rangle)^2~.
\label{ml7}
\ee
If we choose $\langle \phi \rangle /\langle \phi^\prime \rangle \approx 10$, and then set
$\langle \phi \rangle \approx 100$ GeV and  $M_X \sim 10-100$ TeV, 
we must require $f^\prime \leq 10^{-4}-0.1$ in order to satisfy the current limit 
$M_L \leq  M_{eff}^{exp} = 0.2$ eV. 

But also in this model the condition $M_L, M_R \ll M_D$ is hard to
realize if $\Delta m^2$ is as small as $10^{-10}$ eV which corresponds
to the vacuum oscillation solution for the solar neutrino problem. For $M_X=10$ TeV ($M_X=100$ TeV), 
$f^\prime$ must have unnaturally small values, $f^\prime < 10^{-9}$
($f^\prime < 10^{-6}$).

From this example one can conclude that in the 
pseudo-Dirac scenario the scale of new physics could be very high and
that $M_L$ and $M_D$ are outside laboratory detection at present and
also for any future realistic experiments. Naturally, it 
follows that, if for example
GENIUS observes a nonzero signal for the $0\nu\beta\beta$ decay,
pseudo-Dirac scenario is very unlikely. 

\bigskip 
 
\noindent{\it Cosmological constraints.} Here, we discuss the constraints
assuming that the new physics arises from an operator of the type 
${\cal L}_5$ and is consistent with pseudo-Dirac scenario. In the context 
of cosmology, neutrinos being neutral can be ideal candidates for the hot
 dark matter. In the non-relativistic limit, the energy density is
 $\rho_\nu = \sum_i  
m_{\nu_i}N_\nu$, where $N_\nu$ is the number density and $m_{\nu_i}$ 
are the mass values. In the context of four neutrino flavours, it is expected 
that there is at least a pair of nearly degenerate neutrinos. It is possible
 that the splitting between such nearly degenerate pairs could correspond to 
the solar sector. It is conceivable that the dark matter is 
composed of some fraction of such degenerate or nearly degenerate neutrinos 
with the splitting to be 
$\sim \sqrt{\Delta m^2_\odot} \approx 10^{-4} - 10^{-5}$ eV; this value 
of the mass splitting in our case will be close to the Dirac mass. 
Therefore, for such quasi-degenerate
masses  $m_{\nu}\approx M_D$, we can relate to the cosmological parameters
 as \cite{cosmo} 
\be 
\sum_\alpha M_D \approx 94\,\Omega_{\nu}\,\textrm{eV}~, 
\label{eq10} 
\ee 
\noindent 
where $\Omega_{\nu}$ is the neutrino density compared with the
critical density, and $\alpha$ runs from 1 to $n_f$, 
where $n_f$ is the number of flavors in thermal equilibrium.  
Using (\ref{eq3.1}) and (\ref{j5}), we can rewrite (\ref{eq10}) as 
 
\be 
\Delta m^2 \approx 94\,\Omega_{\nu}\beta \frac{M_{eff}}{n_f}~{\mbox{eV}}.
\label{eq11} 
\ee 
\noindent 
The present allowed range is 
$0.003 <\Omega_{\nu}< 0.1~$ \cite{kappa}. This yields the lower limit 
 
\be  
\frac{0.1\,n_f\Delta m^2}{\beta\,\mbox{eV}}\lsim M_{eff}~. 
\ee           
 
Comparing this with the lower limit for $M_{eff}$ in~(\ref{mlbound}),  
which was obtained by requiring that the scale $M_X \leq M_{pl}$,  
one notices that the bound obtained from cosmology is more stringent only if   
\be  
\Delta m^2 n_f \gsim 4.7 \cdot 10^{-5}~f\beta~\mbox{eV}^2~.
\label{c1}   
\ee 
This is not in accordance with the vacuum oscillation solution of 
the solar deficit problem which requires $\Delta m^2 \sim 10^{-10}$ 
eV$^2$. Therefore we conclude that in the limit of the dark matter being composed of some fraction
of degenerate neutrinos, cosmology does not give more stringent  
bounds on $\Delta m^2$ than the oscillation results.

\section{summary} 
 
We have investigated a pseudo-Dirac mixing of left and right-handed neutrinos 
assuming that the 
Majorana masses $M_L$ and $M_R$ are small compared with the Dirac mass $M_D$. 
In this scenario there exist light sterile neutrinos, which may be necessary 
for explaining the solar 
and atmospheric neutrino anomalies together with the LSND results on neutrino 
oscillations. We assume
 that the Majorana mass $M_L$ is generated by a non-renormalizable operator 
reflecting effects of new 
physics at some high energy scale. A consistent scenario relevant for the pure 
$\nu_e \leftrightarrow \nu_s$
vacuum oscillation is achieved with $M_L,M_R\simeq 10^{-7}$ eV and 
$M_D\simeq 10^{-5}- 10^{-4}$ eV. In this case, the preferred value for 
$M_D$ is pushed to its upper
end which arises due to  the pseudo-Dirac criterion $(M_D\gg
M_L)$. The mass $M_L$ is easily correlated to the bound for the effective 
Majorana mass as probed in neutrinoless double beta decay searches.  
Unfortunately, the planned future experiments for probing $M_{eff}$
are still (at least) a couple of orders above the required
sensitivity. If on the other hand, future experiments do observe a positive
signal for $M_{eff}$, then this will disfavour a pseudo-Dirac scenario.
If the Majorana mass is to be generated by the simplest non-renormalizable operator (15), then such a positive effect would furthermore imply that the scale of new physics has to be at the GUT scale or otherwise neutrino parameters should be unnaturally fine tuned. An interesting possibility is if the new physics is much below the Planck 
scale along with a nonzero signal for $0\nu\beta\beta$ decay.
In this case, without much fine tuning of the neutrino parameter, the scale 
of new physics could be at the TeV range. This scenario, based on an operator of the form (21), invokes additional scalar doublets with
a possible $Z_2$ symmetry. However, also in this case the pseudo-Dirac scenario were ruled out.
We also show that in the limit of nearly degenerate neutrino as dark 
matter components, the corresponding bounds for the neutrino
parameters are less stringent than the ones obtained due to
oscillations. This is primarily due to the small mass squared
difference required for the solar solution.

\begin{center} 
{\bf Acknowledgments} 
\end{center} 
 
  Balaji wishes to thank the hospitality at the Theoretical 
Physics Division, University of Helsinki, Finland, where this work 
was started. We also thank E.A. Paschos and Werner Rodejohann for useful 
comments. This work has been supported by the Academy of Finland 
under the contract no. 40677 and also by the
Bundesministerium f\"ur Bildung, Wissenschaft, Forschung und Technologie,
Bonn under contract no. 05HT1PEA9.

\end{document}